\documentclass{article}
\usepackage{spconf,amsmath,graphicx,hyperref}
\usepackage{amssymb}
\usepackage{booktabs}
\usepackage{CJK}

\title{Steer-MoE: Efficient Audio-Language Alignment with a Mixture-of-Experts Steering Module}
\name{Ruitao Feng$^{1}$ \qquad Bixi Zhang$^{2}$ \qquad Sheng Liang$^{1}$ \qquad Zheng Yuan$^{\star}$ }

\address{$^{1}$ Independent Researcher, China \\
    $^{2}$ The University of Hong Kong, Fauclty of Science, Hong Kong \\
    $^{\star}$ Aix-Marseille University, Laboratoire Parole et Langage (LPL), France}

\begin{document}
%
\maketitle
\begin{abstract}
Aligning pretrained audio encoders and Large Language Models (LLMs) offers a promising, parameter-efficient path to building powerful multimodal agents. However, existing methods often require costly full-model finetuning or rely on static adapters that may lack expressive power. Drawing inspiration from the Platonic Representation Hypothesis, we introduce SteerMoE, a novel and modular framework for audio-language alignment. SteerMoE freezes both the audio encoder and the LLM decoder, training only a lightweight steering module integrated within the encoder's layers. This module uses a Mixture-of-Experts (MoE) router to dynamically select and apply learned steering vectors, progressively transforming continuous audio representations into a space comprehensible to the LLM. By operating entirely in the continuous embedding space, our approach requires no modifications to the LLM's vocabulary and preserves its advanced reasoning and agentic capabilities. We demonstrate through experiments on ASR, audio understanding, and a qualitative function-calling task that SteerMoE achieves strong performance while remaining highly modular and computationally efficient, offering a robust new paradigm for developing sophisticated audio-language systems.

\end{abstract}


\section{Introduction}
\label{sec:intro}
The human brain seamlessly integrates rich sensory inputs---sight, sound, and language---into a coherent model of the world. A central goal in artificial intelligence is to imbue machines with a similar capability, enabling them to reason across diverse data modalities. Recent theoretical work has formalized this pursuit under the Platonic Representation Hypothesis~\cite{Huh-2024-platonic}, which posits that neural networks, when trained on varied data, converge towards a shared, underlying representation of reality, much like Plato's Forms. According to this hypothesis, the projections of this universal reality onto different modalities (e.g., the visual appearance of a cat versus the sound of its meow) can be reconciled through simple, interpretable transformations~\cite{huang-etal-2025-cross, wu-2023-large,tjandrasuwita2025understanding}.

This perspective offers a powerful paradigm for multimodal AI, suggesting that bridging the gap between modalities may not require monolithic, end-to-end training, but rather the discovery of efficient alignment functions within this shared latent space. In this work, we investigate this hypothesis within the audio-language domain, a critical nexus for human-computer interaction. We explore the challenge of aligning the continuous world of acoustics with the symbolic world of large language models (LLMs). Our contribution is a novel, parameter-efficient architecture that achieves this alignment through dynamic, context-aware steering, demonstrating a modular and effective path towards multimodal AI agents.

Current efforts to build audio-language models predominantly fall into two categories. The first involves training large, monolithic models from scratch or through extensive end-to-end finetuning on massive audio-text corpora~\cite{chu-2024-qwen2, kimiteam-2025-kimi, xu-2025-fireredasr, wang-2025-teaching}. While these models achieve state-of-the-art performance, their development demands prohibitive computational resources, results in highly-coupled, inflexible architectures, and risks degrading the LLM's original reasoning capabilities. The second, more efficient paradigm involves parameter-efficient finetuning (PEFT), where a pretrained audio encoder is adapted to a frozen LLM. A common PEFT strategy is to discretize audio into a sequence of acoustic tokens, expanding the LLM's vocabulary to treat audio as another "language"~\cite{shaik-2024-lara, kim-2024-para}. This, however, introduces architectural complexity via a separate quantizer, risks information loss, and compromises the LLM's modularity. A more direct approach uses a static adapter, such as a simple MLP, to map continuous audio features into the LLM's embedding space as a soft prompt~\cite{Sammani-2025-zero, hu-2024-wavllm}. While parameter-efficient, such static mappings may lack the expressive power to perform the nuanced, context-dependent alignment required for diverse and complex speech tasks.

In this paper, we introduce SteerMoE, a framework that navigates a middle path between costly end-to-end training and overly simplistic adaptation. Our approach directly operationalizes the Platonic Representation Hypothesis by learning a dynamic, content-aware alignment function. We insert a lightweight Mixture-of-Experts (MoE) module that operates internally within a frozen audio encoder. This module learns to select and apply a combination of expert "steering vectors" to the audio representations at each layer, adaptively modifying them to be seamlessly understood by a frozen LLM decoder. By manipulating representations directly in the continuous vector space, SteerMoE entirely bypasses audio tokenization, preserving the full richness of the acoustic signal and leaving the LLM's architecture untouched. This creates a truly 'plug-and-play' framework where components can be interchanged with ease. Our main contributions are:
\begin{itemize}
    \item We introduce a dynamic, layer-wise steering mechanism based on MoE, offering a more expressive and parameter-efficient alignment than static adapters.
    \item We present a fully modular framework where audio encoders (e.g., Whisper, Conformer) and LLM decoders (e.g., Qwen, LLaMA) can be independently swapped, preserving their native reasoning and agentic capabilities.
    \item We provide strong experimental evidence that such a lightweight steering approach is sufficient to align audio and language representations by training and evaluating our model on diverse tasks, including Automatic Speech Recognition (ASR) on LibriSpeech and AISHELL-2, and audio understanding on Clotho-AQA benchmark. 
\end{itemize}

\section{Methodology}
\label{sec:methodology}

\begin{figure}[h]
    \centering
    \includegraphics[width=0.9\columnwidth]{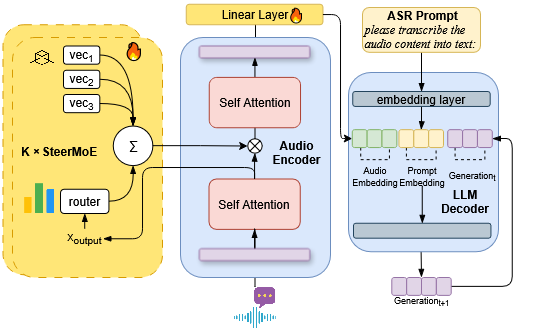}
    \caption{The SteerMoE architecture. A frozen audio encoder processes the input waveform. At each layer, a trainable MoE steering module refines the audio representations. The final steered features are projected and used as a continuous prompt for a frozen LLM decoder.}
    \label{fig:architecture}
\end{figure}

Our proposed framework, SteerMoE, is designed to be a modular and parameter-efficient solution for aligning audio and language representations. It consists of three primary components: a frozen pretrained audio encoder, a frozen pretrained LLM decoder, and a lightweight, trainable steering module that operates within the encoder's layers. The overall architecture is depicted in Figure~\ref{fig:architecture}.

\subsection{Efficient Layer-wise Steering Module}
\label{ssec:steering_module}

The core of our method is a dynamic, layer-wise steering module that progressively refines the audio representations within the audio encoder. For an encoder with $L$ layers, the module applies a content-aware adjustment at each layer. This module comprises a set of expert steering vectors, a shared router, a linear projection layer, and learnable scaling factors.

For each encoder layer $l \in \{1, ..., L\}$, we define a set of $N$ learnable expert steering vectors, $\{E_{l,n}\}_{n=1}^{N}$, where each $E_{l,n} \in \mathbb{R}^{D}$ and $D$ is the feature dimension of the encoder.

To maintain parameter efficiency, a single, shared MoE router, implemented as a linear layer with weights $W_{router} \in \mathbb{R}^{D \times (L \cdot N)}$, generates gating logits for all experts across all layers. Given the hidden state sequence $H_l \in \mathbb{R}^{T \times D}$ from the output of layer $l$, where $T$ is the sequence length, the router computes gating scores $g_l \in \mathbb{R}^{T \times N}$ for that layer's experts as follows:
\begin{equation}
    g_l = \text{softmax}(\text{slice}_l(H_l W_{router}))
    \label{eq:gating}
\end{equation}
where $\text{slice}_l(\cdot)$ is an operation that extracts the $N$ logits corresponding to layer $l$. The resulting steering adjustment $\Delta H_l$ is the weighted sum of the expert vectors for that layer:
\begin{equation}
    \Delta H_l = g_l E_l
    \label{eq:adjustment}
\end{equation}
This adjustment is then scaled by a learnable, per-layer parameter $\alpha_l$ and added back to the original hidden state to produce the final steered output $H'_l$:
\begin{equation}
    H'_l = H_l + \alpha_l \Delta H_l
    \label{eq:steered_output}
\end{equation}

After the final steering operation at the $L$-th layer, we apply an average pooling layer with a kernel size of 4 across the temporal dimension. This downsampling step reduces the sequence length of the audio features, enhancing computational efficiency before they are passed to the LLM decoder. The resulting steered audio representation, $H'_{\text{audio}}$, is then interfaced with the LLM.

\subsection{Modality Alignment via Continuous Prompting}
\label{ssec:prompting}
Our framework interfaces the steered audio representations with the LLM decoder by treating them as a continuous, or "soft", prompt. This approach operates entirely in the continuous vector space, avoiding the information loss and architectural modifications associated with discrete audio tokenization.

The final sequence of steered audio vectors from the encoder, $H'_{audio} \in \mathbb{R}^{T_{audio} \times D}$, is first passed through a trainable linear projection layer with weights $W_{proj} \in \mathbb{R}^{D \times D_{llm}}$ to match the LLM's hidden dimension, $D_{llm}$. This results in a sequence of audio prompt embeddings $P_{audio}$. These embeddings are then prepended to the standard text embeddings $E_{text}$ to form the final input sequence for the LLM decoder:
\begin{equation}
    E_{input} = [H'_{audio} W_{proj} ; E_{text}] = [P_{audio} ; E_{text}]
    \label{eq:concatenation}
\end{equation}

\subsection{Training Objective}
\label{ssec:objective}

The trainable parameters of SteerMoE---comprising the expert steering vectors $\{E_{l,n}\}$, the shared router $W_{router}$, the scaling factors $\{\alpha_l\}$, and the projection layer $W_{proj}$---are optimized using a standard auto-regressive, next-token prediction objective. The model learns to predict the next token in the text sequence, conditioned on the audio prompt and the preceding text tokens.

Crucially, the cross-entropy loss is computed \textit{only} over the target text tokens. The audio prompt portion of the input sequence is masked out from the loss calculation. Given a target text sequence $Y = \{y_1, ..., y_{T_{text}}\}$, the objective is to minimize the negative log-likelihood:
\begin{equation}
    \mathcal{L}(\theta) = -\sum_{t=1}^{T_{text}} \log p(y_t | P_{audio}, y_{<t}; \theta)
    \label{eq:loss}
\end{equation}
where $\theta$ represents all trainable parameters. This ensures that the steering module learns to transform audio representations into a format that the frozen LLM can effectively use to condition its text generation.

\section{Experiments}
\label{sec:experiments}

We evaluate our SteerMoE framework on two distinct task categories: foundational \textbf{Automatic Speech Recognition (ASR)} and complex \textbf{Audio Question Understanding (AQU)}.

\subsection{Datasets and Tasks}
\label{ssec:datasets}

To measure the core quality of the audio-to-text alignment, we use two standard ASR benchmarks:
\begin{itemize}
    \item \textbf{LibriSpeech}~\cite{panayotov-2015-librispeech}: A public domain corpus for English speech recognition, derived from audiobooks.
    \item \textbf{AISHELL-2}~\cite{du-2018-aishell}: A 1000-hour corpus for Mandarin Chinese ASR, recorded in a quiet indoor environment.
\end{itemize}

To evaluate the model's ability to perform complex reasoning over audio, we use the Clotho-AQA benchmark.

\begin{itemize}
    \item \textbf{Clotho-AQA}~\cite{Lipping-2022-clotho} is a publicly available audio question answering dataset consisting of 1,991 audio samples of 15–30 seconds each, drawn from the Clotho dataset. Each sample has six questions, and for each question, three different annotators provide answers, yielding a total of 35,838 question-answer pairs. Four questions per sample are binary (yes/no) and two are single-word answer questions. 
    
\end{itemize}

\subsection{Implementation Details}
\label{ssec:implementation}

Our SteerMoE model is built upon powerful, publicly available pretrained components. We use the Whisper-large-v3 model as our frozen audio encoder and the Qwen2.5-7B-Instruct model as our frozen LLM decoder. The only trainable parameters are those within our steering module and the final projection layer. The steering module employs $N=8$ experts with an initial steering scale $\alpha_l=0.1$. Training is performed with a batch size of 4, AdamW optimizer, FP16 precision, and separate learning rates for the base model ($1\times10^{-4}$), steering vectors ($1\times10^{-2}$), and router ($1\times10^{-3}$).

\begin{CJK}{UTF8}{gbsn}
Audio inputs are standardized to 16kHz mono, and log-Mel spectrogram features are extracted using the Whisper feature extractor. Text transcriptions are tokenized with the Qwen tokenizer, prepended with an instructional prompt: ``\textit{please transcribe the audio content into text: }'' for English ASR task, ``\textit{请逐字复述音频内容为文字}''  (Please transcribe the audio input word by word) for Chinese ASR task, and ``\textit{lease answer the following question. The question is : }'' + batch["Question Text"] for audio understanding task. We filter samples longer than 30 seconds or 448 text tokens due to the limitation of computational resources. A custom data collator handles padding and masks the audio prompt tokens from the loss calculation.
\end{CJK}

\subsection{Main Results}
\label{ssec:main_results}

We compare SteerMoE against strong published results for representative baseline models on both ASR and AQU tasks.

\subsubsection{ASR Performance}
As shown in Table~\ref{tab:asr_results}, our parameter-efficient SteerMoE model achieves competitive ASR performance. To explicitly validate the ``plug-and-play'' modularity of our approach, we tested two different frozen audio encoders: Whisper-large-v3 and Conformer. While not yet matching the performance of a fully-finetuned system, both configurations demonstrate strong transcription capabilities by training only a minuscule fraction of the total parameters, validating the effectiveness of our alignment strategy.

\begin{table}[h]
\centering
\caption{ASR results (CER/WER \%) on LibriSpeech and AISHELL-2 test sets. Lower is better. W and C in the model name indicate the Whisper-large-v3 or Conformer encoder applied and the number after its parameter size}
\label{tab:asr_results}
\begin{tabular}{lcc}
\toprule
\textbf{Model} & \textbf{LS (WER)} & \textbf{AS-2 (CER)} \\
\midrule
Whisper-large-v3 & 2.7 & 4.96 \\
SteerMoE (W7B) & 5.69 & 5.96 \\
SteerMoE (C3B) & 3.26 & 3.44 \\
\textbf{SteerMoE (C7B)} & \textbf{2.42} & \textbf{2.50} \\
\bottomrule
\end{tabular}
\end{table}

\subsubsection{Audio Understanding Performance}
On the challenging ClothoAQA benchmark, SteerMoE demonstrates its ability to unlock the LLM's powerful reasoning capabilities for spoken inputs. As shown in Table~\ref{tab:sqa_results}, our model performs favorably against strong, larger multimodal models, highlighting the benefits of preserving the LLM's integrity through our frozen-decoder approach.

\begin{table}[h]
\centering
\caption{Performance on the ClothoAQA benchmark (Average Accuracy \%). Higher is better. Includes both industrial multimodal LLMs and our SteerMoE models. Data for the industrial models taken from \url{https://github.com/MoonshotAI/Kimi-Audio}.}
\label{tab:sqa_results}
\begin{tabular}{lcc}
\toprule
\textbf{Model} & \textbf{Param.} & \textbf{Avg. Acc. (\%) }\\
\midrule
Kimi-Audio & 9.77B & 71.24 \\
Step-Audio-Chat & 130B & 45.84 \\
\midrule
\multicolumn{3}{c}{\textbf{}} \\
\midrule
\textbf{SteerMoE (W7B)} & \textbf{7B + 1.5B + 64M} & \textbf{52.35} \\
SteerMoE (C3B) & 3B + 1.5B + 64M & 46.24 \\
SteerMoE (C7B) & 7B + 1.5B + 64M & 49.06 \\
\bottomrule
\end{tabular}
\end{table}

\subsubsection{Qualitative Analysis of Agentic Capabilities}
\begin{CJK}{UTF8}{gbsn}
A key claim of our work is that by freezing the LLM, its native capabilities are preserved. To test this, we conducted a simple qualitative experiment to probe the model's agentic function-calling ability. We configured a tool-use scenario where a specific spoken query, ``\textit{上海的天气怎么样?}'' (What's the weather in Shanghai?), should trigger a predefined function that returns a non-literal, absurd answer. When presented with the audio of this query, our SteerMoE model correctly interpreted the user's intent from speech and successfully triggered the function call, yielding the predefined response: ``\textit{上海今天下黄金}'' (It's raining gold in Shanghai today). This successful outcome demonstrates that the complex machinery for tool-use and function calling, inherent to the frozen Qwen LLM, remains fully intact and is directly accessible via spoken commands through our alignment module.
\end{CJK}

\subsection{Ablation Studies}
\label{ssec:ablation}

To validate our architectural design, we conducted several ablation studies, with results summarized in Table~\ref{tab:ablation_results}. First, we replaced our dynamic MoE-based steering module with a simple linear projection layer (a static adapter). The significant performance drop underscores the importance of a dynamic, content-aware alignment mechanism. Second, we experimented with varying the number of expert vectors per layer, finding that using 8 experts provides a strong balance of performance and parameter efficiency compared to 2 or 4 experts. These results confirm that the expressiveness of the MoE router is a critical component of our model's success.

\begin{table}[h]
\centering
\caption{Ablation studies on LibriSpeech.}
\label{tab:ablation_results}
\begin{tabular}{lc}
\toprule
\textbf{Model Variant} & \textbf{WER (\%) } \\
\midrule
\textbf{SteerMoE (8 Experts)} & \textbf{2.42\%} \\
SteerMoE (4 Experts) & 3.10\% \\
SteerMoE (2 Experts) & 6.22\% \\
Static Adapter (No MoE) & 103\% \\
\bottomrule
\end{tabular}
\end{table}

\section{Discussion}
\label{sec:discussion}

Our results demonstrate that a lightweight, dynamic steering module can effectively align the representational spaces of separate, pretrained audio and language models. The success of this parameter-efficient approach, which operates entirely in the continuous embedding space, provides strong empirical support for the Platonic Representation Hypothesis in the audio-language domain. It suggests that complex, deep fusion is not a prerequisite for multimodal understanding; rather, discovering the correct transformations within a shared conceptual space is sufficient.

The primary advantage of our method is its modularity and preservation of the LLM's inherent capabilities. By freezing the decoder and avoiding vocabulary modification, SteerMoE makes the core components ``plug-and-play'' and, more importantly, unlocks the LLM's advanced reasoning and agentic abilities for spoken input, as shown in our function-calling experiments. This positions our work not just as an ASR system, but as a general framework for building sophisticated audio-language agents.

Nonetheless, we acknowledge several limitations. Our experiments were conducted on relatively clean speech corpora. The model's robustness in highly noisy environments and its performance on a wider diversity of non-speech audio remain open questions. Furthermore, while our training was constrained by data scale and sequence length due to available computational resources, our architectural design theoretically allows the model at inference to process audio sequences of any length that the underlying LLM decoder's context window can accommodate. Our agentic experiment, while a successful proof-of-concept, also requires testing on more complex, multi-turn interactions.

For future work, a particularly exciting direction is to probe the learned expert steering vectors. Analyzing whether specific experts come to specialize in phonetic features, prosody, speaker identity, or even noise separation could yield fascinating insights into the nature of the learned audio-language mapping.

\section{Compliance with Ethical Standards}
\label{ethic}

This research study was conducted retrospectively using only human-subject data made available in open access by the respective dataset providers stated in the original papers cited (LibriSpeech, AISHELL-2, and Clotho-AQA). Ethical approval was not required as confirmed by the licenses attached with the open access data.

\bibliographystyle{IEEEbib}
\bibliography{refs}

\end{document}